\pdfoutput=1

\makeatletter
\def\cl@chapter{}
\makeatother

\documentclass[twocolumn,natbib,english]{svjour3}        

\makeatletter
\if@twocolumn
  \renewcommand\normalsize{%
    \@setfontsize\normalsize\@xpt{12.5pt}%
    \abovedisplayskip=3 mm plus6pt minus 4pt
    \belowdisplayskip=3 mm plus6pt minus 4pt
    \abovedisplayshortskip=0.0 mm plus6pt
    \belowdisplayshortskip=2 mm plus4pt minus 4pt
    \let\@listi\@listI}%

  \renewcommand\small{%
    \@setfontsize\small{8.5pt}\@xpt
    \abovedisplayskip 8.5\p@ \@plus3\p@ \@minus4\p@
    \abovedisplayshortskip \z@ \@plus2\p@
    \belowdisplayshortskip 4\p@ \@plus2\p@ \@minus2\p@
    \def\@listi{\leftmargin\leftmargini
      \parsep 0\p@ \@plus1\p@ \@minus\p@
      \topsep 4\p@ \@plus2\p@ \@minus4\p@
      \itemsep0\p@}%
    \belowdisplayskip \abovedisplayskip}
\else
  \if@smallext
    \renewcommand\normalsize{%
      \@setfontsize\normalsize\@xpt\@xiipt
      \abovedisplayskip=3 mm plus6pt minus 4pt
      \belowdisplayskip=3 mm plus6pt minus 4pt
      \abovedisplayshortskip=0.0 mm plus6pt
      \belowdisplayshortskip=2 mm plus4pt minus 4pt
      \let\@listi\@listI}%

    \renewcommand\small{%
      \@setfontsize\small\@viiipt{9.5pt}%
      \abovedisplayskip 8.5\p@ \@plus3\p@ \@minus4\p@
      \abovedisplayshortskip \z@ \@plus2\p@
      \belowdisplayshortskip 4\p@ \@plus2\p@ \@minus2\p@
      \def\@listi{\leftmargin\leftmargini
        \parsep 0\p@ \@plus1\p@ \@minus\p@
        \topsep 4\p@ \@plus2\p@ \@minus4\p@
        \itemsep0\p@}%
      \belowdisplayskip \abovedisplayskip}
  \else
    \renewcommand\normalsize{%
      \@setfontsize\normalsize{9.5pt}{11.5pt}%
      \abovedisplayskip=3 mm plus6pt minus 4pt
      \belowdisplayskip=3 mm plus6pt minus 4pt
      \abovedisplayshortskip=0.0 mm plus6pt
      \belowdisplayshortskip=2 mm plus4pt minus 4pt
      \let\@listi\@listI}%

    \renewcommand\small{%
      \@setfontsize\small\@viiipt{9.25pt}%
      \abovedisplayskip 8.5\p@ \@plus3\p@ \@minus4\p@
      \abovedisplayshortskip \z@ \@plus2\p@
      \belowdisplayshortskip 4\p@ \@plus2\p@ \@minus2\p@
      \def\@listi{\leftmargin\leftmargini
        \parsep 0\p@ \@plus1\p@ \@minus\p@
        \topsep 4\p@ \@plus2\p@ \@minus4\p@
        \itemsep0\p@}%
      \belowdisplayskip \abovedisplayskip}
  \fi
\fi
\let\footnotesize\small
\makeatother

\usepackage{cmap}

\RequirePackage[%
  shrink=40,
  final,%
  expansion=alltext,%
  protrusion=alltext-nott]{microtype}%
\DisableLigatures{encoding = T1, family = tt* }

\usepackage[english]{babel}

\usepackage{ifluatex}
\ifluatex
  \usepackage{fontspec}
  \usepackage[english]{selnolig}
\fi

\ifluatex
\else
  \usepackage[%
    rm={oldstyle=false,proportional=true},%
    sf={oldstyle=false,proportional=true},%
    tt={oldstyle=false,proportional=true,variable=false},%
    qt=false%
  ]{cfr-lm}
\fi

\ifluatex
\else
  \usepackage[T1]{fontenc}
  \usepackage[utf8]{inputenc} 
\fi

\smartqed  

\usepackage{graphicx}

\usepackage{upquote}

\usepackage{booktabs}

\usepackage{paralist}

\usepackage[inline]{enumitem}

\usepackage{csquotes}

\usepackage{textcmds}

\usepackage{url}
\makeatletter
\g@addto@macro{\UrlBreaks}{\UrlOrds}
\makeatother

\usepackage[mode=text,group-four-digits,group-separator={,}]{siunitx}
\usepackage{xcolor}

\usepackage{listings}
\renewcommand{\lstlistingname}{List.}
\usepackage{chngcntr}
\AtBeginDocument{\counterwithout{lstlisting}{section}}

\usepackage{pdfcomment}
\newcommand*{\doi}[1]{\href{https://doi.org/\detokenize{#1}}{DOI: \detokenize{#1}}}

\usepackage{stfloats}
\fnbelowfloat

\usepackage[all]{hypcap}

\usepackage[acronym,indexonlyfirst,nomain]{glossaries}
\glsdisablehyper

\usepackage[capitalise,nameinlink]{cleveref}
\crefname{section}{Sect.}{Sect.}
\Crefname{section}{Section}{Sections}
\crefname{listing}{\lstlistingname}{\lstlistingname}
\Crefname{listing}{Listing}{Listings}
\crefname{define}{Def.}{Def.}
\Crefname{define}{Definition}{Definition}

\usepackage{xspace}
\newcommand{\eg}{e.\,g.,\ }
\newcommand{\ie}{i.\,e.,\ }

\DeclareFontFamily{U}{MnSymbolC}{}
\DeclareSymbolFont{MnSyC}{U}{MnSymbolC}{m}{n}
\DeclareFontShape{U}{MnSymbolC}{m}{n}{
  <-6>    MnSymbolC5
  <6-7>   MnSymbolC6
  <7-8>   MnSymbolC7
  <8-9>   MnSymbolC8
  <9-10>  MnSymbolC9
  <10-12> MnSymbolC10
  <12->   MnSymbolC12%
}{}
\DeclareMathSymbol{\powerset}{\mathord}{MnSyC}{180}

\renewcommand{\eg}{e.\,g.,\xspace}

\hypersetup{hidelinks,
  colorlinks=true,
  allcolors=black,
  pdfstartview=Fit,
  breaklinks=true}

\journalname{JOURNALNAME}

\usepackage[math]{blindtext}
\usepackage{mwe}

\usepackage[amsmath,hyperref]{ntheorem}
\makeatletter
\renewtheoremstyle{plain}
  {\item{\theorem@headerfont ##1\ ##2\theorem@separator}~}
  {\item{\theorem@headerfont ##1\ ##2\ (##3)\theorem@separator}~}
\makeatother

{\theoremheaderfont{\upshape\bfseries}
 \theorembodyfont{\normalfont\slshape}
 \newtheorem{define}{Definition}}

\usepackage{footnote}
\makesavenoteenv{tabular}
\makesavenoteenv{tabular*}
\makesavenoteenv{table}
\makesavenoteenv{table*}

\crefformat{footnote}{#2\footnotemark[#1]#3}

\usepackage{draftwatermark}
\SetWatermarkText{preprint}
\SetWatermarkScale{1}

\usepackage[IEEE,
  key=Wurster2019_EDMM,
  year=2019,
  publication={SICS Software-Intensive Cyber-Physical Systems},
  doi={10.1007/s00450-019-00412-x},
  doiText={https://doi.org/10.1007/s00450-019-00412-x},
  nocopyright,
  nourl
]{authorarchive}

\begin{document}

\title{The Essential Deployment Metamodel: A Systematic Review of Deployment Automation Technologies
}
\titlerunning{The Essential Deployment Metamodel} 

\author{Michael Wurster \and
  Uwe Breitenb\"{u}cher \and
  Michael Falkenthal \and
  Christoph Krieger \and
  Frank Leymann \and
  Karoline Saatkamp \and
  Jacopo Soldani
}

\institute{M.~Wurster / U.~Breitenb\"{u}cher / M.~Falkenthal / C.~Krieger / F.~Leymann / K.~Saatkamp \at
    Institute of Architecture of Application Systems, University of Stuttgart, Stuttgart, Germany\\
    \email{[lastname]@iaas.uni-stuttgart.de}
    \and
    Jacopo Soldani \at
    Department of Computer Science, University of Pisa, Pisa, Italy\\
    \email{soldani@di.unipi.it}
}

\date{\phantom{Received: date / Accepted: date}}

\maketitle

\begin{abstract}
In recent years, a plethora of deployment technologies evolved, many following a declarative approach to automate the delivery of software components.
Even if such technologies share the same purpose, they differ in features and supported mechanisms.
Thus, it is difficult to compare and select deployment automation technologies as well as to migrate from one technology to another. 
Hence, we present a systematic review of declarative deployment technologies and introduce the Essential Deployment Metamodel (EDMM) by extracting the essential parts that are supported by all these technologies.
Thereby, the EDMM enables a common understanding of declarative deployment models by facilitating the comparison, selection, and migration of technologies.
Moreover, it provides a technology-independent baseline for further deployment automation research.
%
\end{abstract}



\section{Introduction}\label{sec:introduction}

With the advent of DevOps~\citep{Humble2011} as a software development paradigm, the gap between \emph{development} and \emph{operations} is attempted to be eliminated by revising organizational and cultural challenges.
One integral aspect of DevOps is to enable an efficient collaboration by establishing deployment processes that are highly automated~\citep{Humble2010} as manual deployments of services consisting of multiple units is complex, hard to repeat, and error-prone~\citep{Oppenheimer2003}.
Key concepts like \emph{configuration management}~\citep{Delaet2010_ConfigurationTools} and \emph{infrastructure as code}~\citep{Morris2016_infra-as-code} enable a continuous and automated delivery of software components over the entire lifecycle, \eg to install, start, stop, or terminate components.
By describing components and infrastructure of an application in maintainable and reusable \emph{deployment models}, a repeatable end-to-end deployment automation can be established.
Such deployment models can be of \emph{declarative} or \emph{imperative} nature~\citep{Endres2017}:
Declarative models express the desired state into which an application or parts thereof are transferred.
In contrast, imperative models describe the deployment steps in a procedural manner. 
In industry and research, declarative deployment models are widely accepted as the most appropriate approach for application deployment and configuration management~\citep{Herry2011_AutomatedPlanningForConfigurationChanges}.
As a result, a plethora of different technologies have been developed following this approach such as Chef, Puppet, AWS CloudFormation, Terraform, and Kubernetes.

All such technologies aim at automating the deployment of applications, but they differ in supported features and mechanisms. 
For example, Terraform supports the deployment across multiple cloud providers and it is able to target different cloud offerings as a service (XaaS).
Whereas, there are cloud provider-specific technologies, such as AWS CloudFormation, allowing the deployment only on Amazon's cloud services.
Moreover, there are platform-specific technologies, such as Kubernetes, that support only specific \emph{deployment bundles} (container images) or cloud service offerings (\eg restricted to PaaS).
In addition, most technologies use their own modeling language with its own syntax and expressiveness.

As a result, it is difficult to compare technologies by their capabilities as there exists currently no systematic comparison of deployment features and mechanisms. 
Further, as application systems are in constant change, it is challenging to choose an appropriate technology upfront.
Moreover, in cases applications need to be migrated, it is also required to migrate the associated deployment models to the target environment's technology.
This quickly gets cumbersome and requires knowledge about how to translate the features and mechanisms from one provider's technology, \eg AWS, to another one, \eg Azure.
%
In addition, the various technologies currently in use impede systematic deployment automation research as the practical feasibility of new approaches is typically only evaluated using a certain technology.
However, this makes it hard for researchers to understand if proposed approaches can be mapped to their technologies in use.

To tackle these issues, we introduce the Essential Deployment Metamodel (EDMM), which we obtained through a systematic analysis aimed at distilling the essential parts of declarative deployment technologies.
We also show how the analyzed technologies comply semantically with the EDMM and how it can be mapped to native constructs of each technology.
Thereby, the EDMM provides a common denominator of the features of the most important deployment technologies.
This enables (1) a common understanding of declarative deployment technologies. 
Further, it (2) eases the selection of a deployment technology for one's own use case as the EDMM mapping describes how this can be achieved based on a technology-independent model.
In cases where it is required to migrate from one provider to another, it also requires to migrate the associated deployment models.
The EDMM (3) supports and eases such migration processes by knowing the essential elements of deployment technologies.
On top of that, the EDMM facilitates an automated transformation into specific deployment technologies in the context of model-driven architecture (MDA) in order to decide which specific technology to use as late as possible.
%
Finally, our results give researchers (4) the possibility to evaluate their concepts in a technology-agnostic manner by knowing to which technologies an approach can be applied to without disruptive adaptations.
%
The review of technologies revealed that there are general-purpose (GP), provider-specific (ProvS), and platform-specific (PlatS) deployment technologies that enable the description of components, relations, and respective types as main deployment model entities.
Hereafter, \cref{sec:survey} and \cref{sec:categorization} present our review framework and technology classification. 
\cref{sec:edm} defines the EDMM, while \cref{sec:mapping} discuss its mapping to selected technologies. 
Finally, \cref{sec:relatedwork} and \cref{sec:conclusion} discuss related work and draw some concluding remarks. 




\section{Review Framework}\label{sec:survey}

\begin{table}[t]
\scriptsize
\centering
\begin{tabular*}{0.48\textwidth}{l@{\extracolsep{\fill}}cc}
\toprule
\textbf{Engine} & \textbf{Query} & \textbf{Records} \\ \midrule
ACM &
\begin{tabular}[c]{@{}c@{}}(\texttt{\enquote{infrastructure as code}}\\ \texttt{\enquote{configuration management})}\\ \texttt{AND keywords.author.keyword:}\\\texttt{(\enquote{cloud computing})}\footnote{Used \enquote{Any field} and \enquote{Matches any} operators.}\end{tabular} &
19 \\ \midrule
IEEE &
\begin{tabular}[c]{@{}c@{}}((\texttt{\enquote{infrastructure as code} OR} \\\texttt{\enquote{configuration management}) AND}\\
\texttt{\enquote{IEEE Terms}:\enquote{cloud computing})}\footnote{\enquote{Command Search} with \enquote{Metadata Only} operator.}\end{tabular} &
71 \\ \midrule
Google &
\begin{tabular}[c]{@{}c@{}}\texttt{\enquote{configuration management}} \\ \texttt{\enquote{infrastructure as code}}\\
\texttt{\enquote{cloud computing}}\end{tabular} &
72.600\footnote{The first 100 results were considered.}\\ \bottomrule
\end{tabular*}
\caption{Search details for identifying deployment technologies.}
\label{tab:search}
\end{table}
%
%
%

This section presents the procedure taken to identify the EDMM.
This was done in three phases:
First, we identified a list of deployment technologies using well-known search engines in research and industry.
Each result was individually considered and searched for presented or used deployment technologies.
In the second phase, we ranked the technologies by the amount of search results in a search engine.
Lastly, the analysis of the highest ranked technologies based on a reference scenario led us to the essential elements of deployment models.
The complete data of the review are available online\footnote{\label{foot:data}\url{http://tinyurl.com/y2azrq3r}}.

\subsection{Phase 1: Identify Technologies}

In the first phase, we used ACM Digital Library, IEEE Xplore, and Google to identify deployment technologies.
We excluded the term \enquote{deployment} as the hundreds of results include unrelated research topics, \ie covering organizational and build processes regarding deployment.
Therefore, we refined the search using known terms to target the identification of deployment automation technologies.
In each search engine, the following query structure was used:
Results must contain the phrase \enquote{infrastructure as code} or \enquote{configuration management} and must match the keyword \enquote{cloud computing}.
The term \enquote{cloud computing} is chosen because we regard the support of cloud services as an elementary characteristic of future-proof deployment technologies.

The exact search queries and result numbers are shown in \cref{tab:search}.
Each result was individually considered and searched for whether a new technology is presented, the technology is used in comparative works, or is used to support evaluating a work or study.
In total, 56 technologies were identified.
We focused our review on open-source or community-licensed technologies and, therefore, excluded eight technologies in this phase since there are only commercial or enterprise licenses available.
Further, we excluded Vagrant~\citep{Vagrant2018} because it focuses on managing local development environments. 
In addition, we excluded AWS OpsWorks~\citep{OpsWorks2018} because it is a managed service by AWS providing Chef or Puppet master nodes and does not provide its own deployment technology.


\subsection{Phase 2: Technology Selection}

To select the most popular deployment technologies, a Google search for each technology was performed.
Since most technologies are industry driven and not backed by scientific papers, the amount of citations is not usable as ranking method.
Therefore, the magnitude of search results in Google was used, which also includes discussions in blogs or newsgroups, in order to derive the relevance of a certain technology.
Even though the result is not precise, we are able to derive a trend in currently used deployment technologies.
We used the following search pattern for ranking:
\enquote{\texttt{<technology> \enquote{configuration management} OR \enquote{infrastructure as code}}}.
We used the American version of Google for the search. 
The search was performed using Google's Chrome browser in incognito mode on February 14, 2019.
We considered the first 13 technologies for further analysis as shown in \cref{tab:ranking}.
The complete ranking is available online\cref{foot:data}.

\subsection{Phase 3: Technology Analysis}\label{subsec:analysis}

We analyzed the features, mechanisms, and capabilities of each selected technology based on a reference scenario, which is depicted in \cref{fig:reference-scenario}.
The authors investigated how it can be realized using the native constructs or extension mechanisms of each technology.

\subsubsection{Deployment Features and Mechanisms}

For the analysis, several aspects were taken into account in order to find commonalities.
We expect the deployment technologies to be 
open-source or community-licensed,
capable to express single-, multi-, or hybrid-cloud deployments, and
provide support for cloud-native application components (PaaS and FaaS).
We derived the following \emph{deployment features and mechanisms} for analyzing the automated deployment capabilities, these are:
(i)~supporting multiple cloud providers and platforms, 
(ii)~targeting different cloud offerings (XaaS), 
(iii)~providing capabilities to structure a deployment into logical parts, 
(iv)~supporting the creation of custom, fine-granular, reusable entities, 
(v)~allowing to specify desired application state, and 
(vi)~allowing to hook into or influence the deployment lifecycle.
These deployment automation features and mechanisms were analyzed based on a reference scenario presented in the next subsection.

\begin{table}[t]
\centering
\begin{tabular*}{0.48\textwidth}{cl@{\extracolsep{\fill}}r}
\toprule
\textbf{\#} & \textbf{Technology} & \textbf{Search Hits} \\ \midrule
1  & Puppet                 & 1.630.000   \\ \midrule
2  & Chef                   & 1.150.000   \\ \midrule
3  & Ansible                & 989.000     \\ \midrule
4  & Kubernetes             & 708.000     \\ \midrule
5  & OpenStack HEAT         & 458.000     \\ \midrule
6  & Terraform              & 348.000     \\ \midrule
7  & AWS CloudFormation     & 156.000     \\ \midrule
8  & SaltStack              & 130.000     \\ \midrule
9  & Juju                   & 53.200      \\ \midrule
10 & CFEngine               & 48.500      \\ \midrule
11 & Azure Resource Manager & 47.100      \\ \midrule
12 & Docker Compose         & 41.300      \\ \midrule
13 & Cloudify               & 23.100      \\ \bottomrule
\end{tabular*}
\caption{Deployment technology ranking by popularity.}
\label{tab:ranking}
\end{table}

\subsubsection{Reference Scenario}

\begin{figure*}[t]
  \centering
  \includegraphics[width=1.0\textwidth]{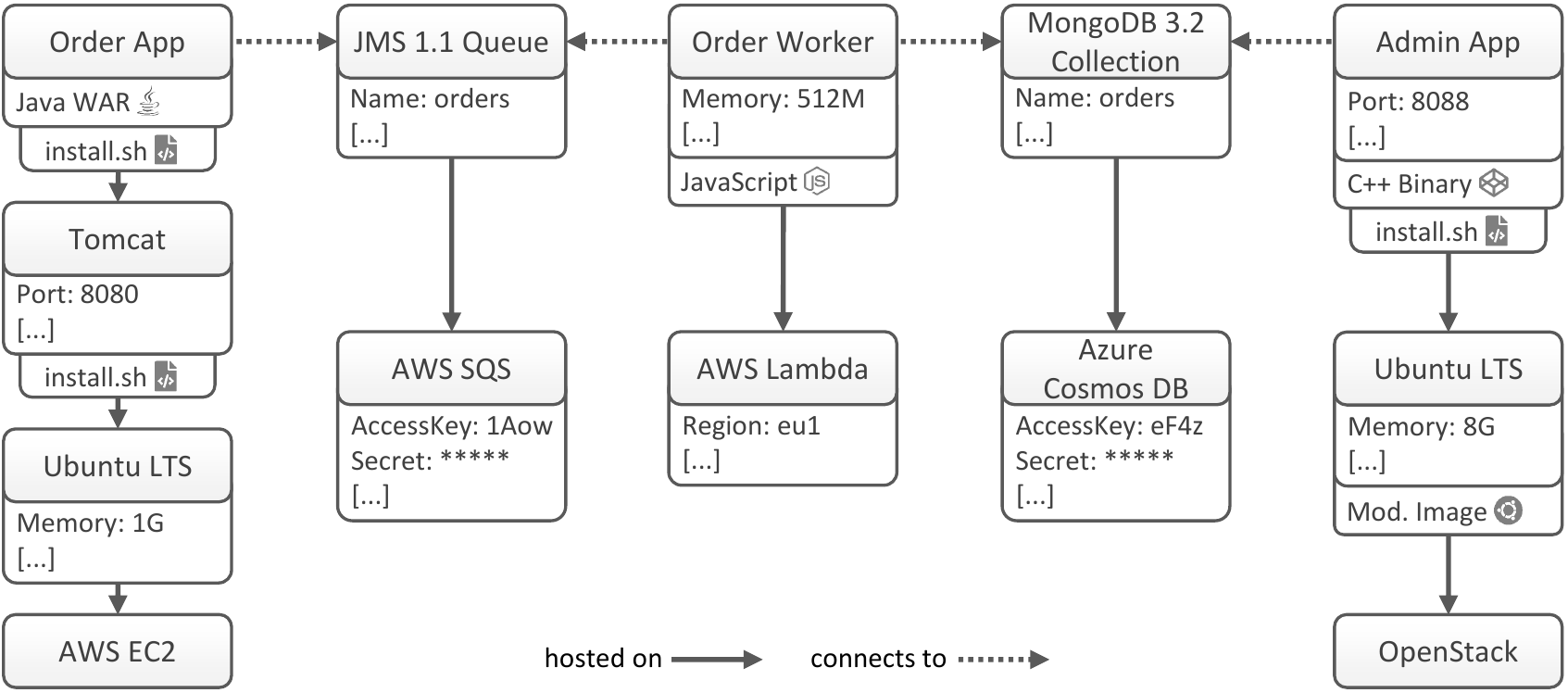}
  \caption{Reference Scenario: Application Topology containing Cloud-native as well as On-Premise Components.}\label{fig:reference-scenario}
\end{figure*}

We envision a simplified multi-cloud order management application containing cloud-native as well as classical and legacy components in order to identify the derived deployment features and mechanisms for each technology.
This reference scenario is intended to cover typical deployment requirements in modern application systems.
It covers exemplary different cloud providers, regarding multi-cloud deployments and different cloud offerings as a service (XaaS). 
Even though the reference scenario does not reflect a complex real-life scenario, it supports the identification of required deployment capabilities to cover modern application deployments.
The left hand side of the figure depicts a Java-based \emph{Order} application deployed onto an Apache Tomcat server, which is depicted to be installed on a virtual machine (VM) provided by Amazon EC2.
The order application is able to push new orders to a queue hosted and managed by Amazon's Simple Queue Service (SQS).
Orders in this queue are processed by a \emph{Order Worker} implemented as an ephemeral and stateless function using Amazon Lambda, Amazon's Function as a Service (FaaS) offering.
In this case, each message put to the queue acts as event and triggers the worker function to process the respective order.
The worker function updates respective values in a database, e.g., allocates the purchase order items for shipping.
To cover a multi-cloud scenario, the orders are stored using a MongoDB Collection hosted and managed by Azure's Cosmos DB service offering.
On the right hand side of the figure, a rather traditional, or non-cloud-native, application component is depicted.
It is envisioned that a kind of administration application is used internally to track and manage the received orders by the \emph{Order App}.
In our scenario, this application is implemented using \texttt{C++}, requiring custom installation routines (cf.~\texttt{install.sh} shell script), and is hosted on-premise using OpenStack. 
In summary, our reference scenario uses \emph{application-specific components}, such as the \emph{Order App} implementing some business logic.
Further, it contains \emph{middleware components}, such as the \emph{Tomcat} server, and \emph{computing components}, \ie virtual machines.
Lastly, it contains so-called \emph{cloud service components}, such as \emph{AWS EC2} and \emph{Azure Cosmos DB}, which are hosting leaf nodes (considering the notion of a graph) and in full control of the respective cloud provider.  
Anyhow, there can be hosting leaf nodes representing traditional VM hypervisors or even bare metal servers. 

\subsubsection{Remarks}

To perform a sound analysis of the selected technologies, we classified the technologies independently.
To ensure avoidance of observer bias, the analysis was executed in parallel over on-third splits of the selected technologies.
Afterwards, the observations and interpretations were discussed and double-checked in several joint sessions for reconciliation.
Based on this analysis and the found commonalities of the technologies, we present a categorization of technologies in the next section.

\pagebreak


\section{Categorization of Technologies}\label{sec:categorization}

During the analysis of the selected technologies (cf.~\cref{tab:ranking}), we observed that they can be divided into three categories. 
Before presenting the categories in detail, we briefly introduce the selected deployment technologies.
Puppet, Chef, and Ansible are configuration management systems.
\textbf{Puppet} enables to write reusable configuration definitions describing system resources and their state for multiple providers and services using their own domain-specific language (DSL)~\citep{Puppet2018}.
%
\textbf{Chef} uses a Ruby-based DSL. 
Based on a server-client architecture, it can be used to maintain and configure systems on various platforms or cloud providers~\citep{Chef2018}.
%
\textbf{Ansible} uses a declarative YAML-based DSL to describe system configurations for various platforms and cloud services.
In contrast to Chef, Ansible uses an \emph{agentless} architecture~\citep{Ansible2018}.
%
\textbf{Kubernetes} is a platform for automating the orchestration of containerized, multi-service applications.
It automatically deploys the specified application onto a cluster and ensures that its desired configuration is reached and maintained~\citep{Kubernetes2018}.
%
\textbf{OpenStack Heat} is an orchestration engine that enables the description of XaaS-based applications using a YAML syntax.
It manages the whole lifecycle of an deployment and provides interfaces for custom extensions~\citep{OpenStackHEAT2018}.
%
\textbf{Terraform} is an orchestrator providing plugin interfaces for custom extensions. 
It uses its own DSL and primarily targets multi-cloud application deployments~\citep{Terraform2018}.
%
In contrast, \textbf{AWS CloudFormation} uses a DSL (JSON and YAML) to describe, deploy, and manage all infrastructure resources across Amazon's cloud services~\citep{Cloudformation2018}.
%
\textbf{SaltStack} is an orchestration and configuration management system using its own DSL to deploy and manage all kinds of application stacks targeting different cloud providers and services~\citep{SaltStack2018}.
%
\textbf{Juju} is a topology-based application orchestration tool.
It enables the modeling of application deployments using a YAML-based DSL and supports multiple cloud offerings and services~\citep{Juju2018}.
%
\textbf{CFEngine} is an open-source configuration management tool providing enterprise functionalities by a commercial version.
Its primary function is to provide automated configuration and management on top of existing computing resources~\citep{Burgess1995}.
%
Similar to AWS CloudFormation, \textbf{Azure Resource Manager} is the deployment and management service by Azure to manage resources in Microsoft's cloud environment~\citep{AzureARM2018}.
%
\textbf{Docker Compose} is a framework for defining and running multi-container Docker applications. 
It provides a YAML-based language to specify the containers forming an application and their configuration~\citep{DockerCompose2018}.
%
\textbf{Cloudify} is an open-source application orchestration framework.
It supports hybrid-cloud deployments for all kinds of cloud services based on a customized YAML syntax inspired by the TOSCA standard~\citep{Cloudify2018}.

\begin{figure}[t]
  \centering
  \includegraphics[width=0.35\textwidth]{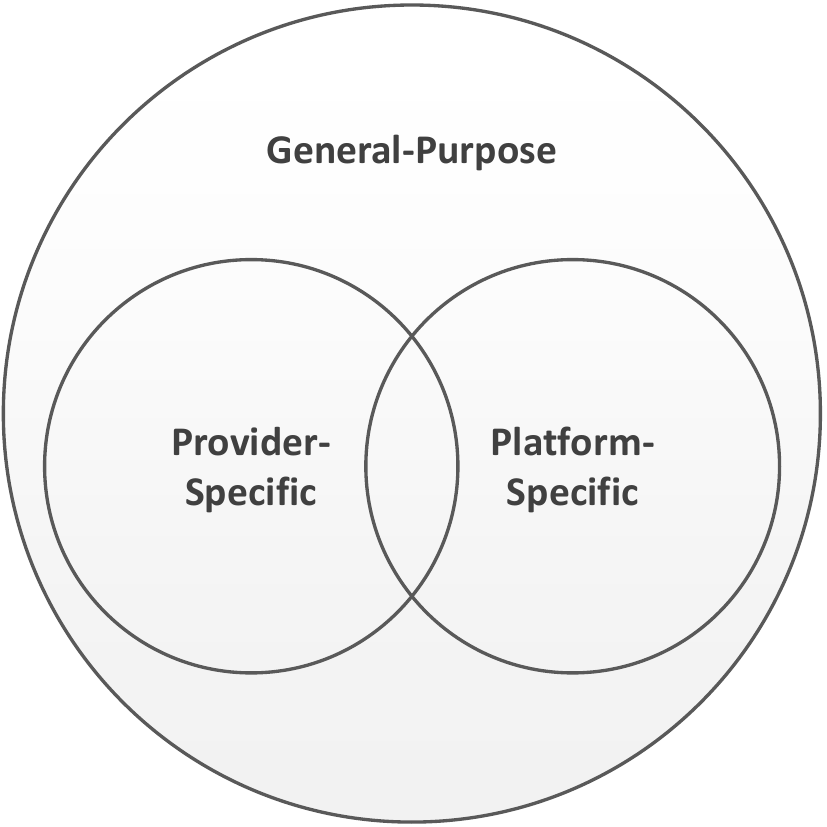}
  \caption{Deployment Automation Technology Categorization}\label{fig:categories}
\end{figure}

During phase three (cf.~\cref{subsec:analysis}), we figured out that some technologies support multi-cloud deployments, such as Ansible or Terraform, and others are restricted to certain cloud providers, such as AWS CloudFormation and Azure Resource Manager. 
Further, we concluded that there are technologies suitable to deploy applications targeting XaaS.
Technologies, such as AWS CloudFormation, Terraform, and Ansible, support different kinds of services for deploying components, whereas technologies, like Kubernetes and Docker Compose, are restricted to use specific \emph{platform bundles}, container images in their cases. 
Therefore, we categorize the technologies in
(i)~general-purpose,
(ii)~provider-specific, and
(iii)~platform-specific deployment technologies.
\Cref{fig:categories} indicates the overlap of deployment technologies regarding the derived \emph{deployment features and mechanisms}.

\begin{description}
\item[\textbf{General-Purpose (GP)}]
GP technologies support all \emph{deployment features and mechanisms} (cf.~\cref{subsec:analysis}).
They support single-, hybrid-, and multi-cloud deployments as well as different kinds of cloud services (XaaS).
In addition, they can be extended by reusable and customized components for further providers or services.
Thereby, it is possible to hook into or influence the component's lifecycle by defining custom actions.
This group encompasses the following technologies:
Puppet, Chef, Ansible, OpenStack Heat, Terraform, SaltStack, Juju, and Cloudify.
\item[\textbf{Provider-Specific (ProvS)}]
ProvS deployment technologies support XaaS deployments, provide capabilities to create reusable entities, and can control a component's lifecycle (cf.~deployment features and mechanisms presented in \cref{subsec:analysis}).
In contrast to GP, ProvS technologies only support single-cloud deployments since they are offered by specific cloud providers, hence only supporting the cloud services offered by the respective provider.
The EDMM is restricted to provider-supported component types for the cloud service components (cf.~\cref{fig:reference-scenario}).
This group encompasses the following deployment technologies:
AWS CloudFormation and Azure Resource Manager.
\item[\textbf{Platform-Specific (PlatS)}]
PlatS deployment techno-logies support multiple cloud providers, the creation of reusable entities, and influencing a component's lifecycle (cf.~deployment features and mechanisms presented in \cref{subsec:analysis}).
In contrast to GP, these are restricted regarding the cloud delivery model and regarding the use of specific \emph{platform bundles} for realizing components.
Considering the reference scenario (cf.~\cref{fig:reference-scenario}), the EDMM is restricted such that only types and artifacts can be used that are supported by the underlying platform, \eg Kubernetes only supports the deployment of container images.
This group encompasses the following deployment technologies:
Kubernetes, CFEngine, and Docker Compose.
\end{description}

\noindent 
All three groups provide deployment features and mechanisms to cover all extracted elements covered by the EDMM presented in the next section.
However, the ProvS and PlatS groups are restricted in the power to express a deployment, which is explained in detail in the technology mapping section (cf.~\cref{sec:mapping}).

\section{The Essential Deployment Metamodel}\label{sec:edm}

\begin{figure*}[t]
  \centering
  \includegraphics[width=0.99\textwidth]{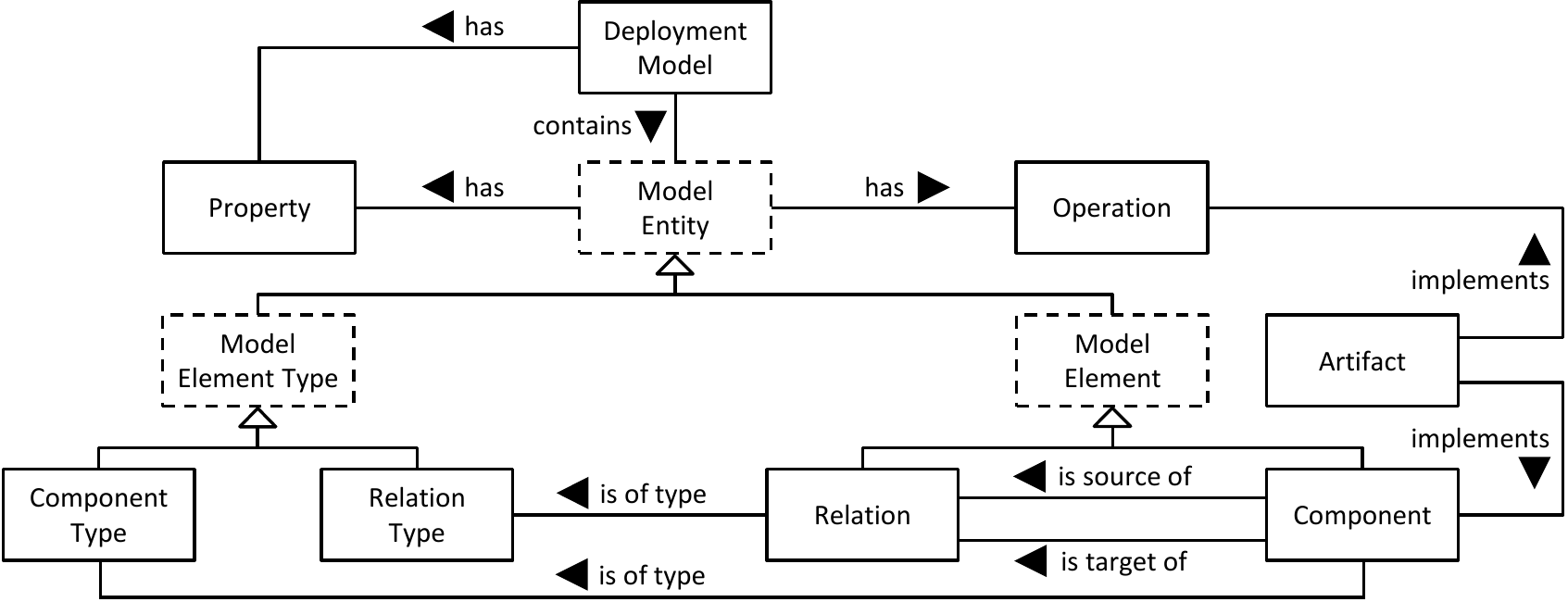}
  \caption{The Essential Deployment Metamodel}\label{fig:metamodel}
\end{figure*}

The EDMM encompasses the essential parts of declarative deployment models.
Declarative deployment models focus on the \enquote{what} and describe the structure of an application to be deployed including all components, their configuration, and relationships.
The EDMM represents the result of the analysis of 13 selected declarative deployment technologies applied in industry and research.
The EDMM is illustrated in \cref{fig:metamodel}, where the structure and names of the entities are inspired by the TOSCA standard~\citep{TOSCA_Specification} and the Declarative Application Management Modeling and Notation (DMMN) with its graph-based nature~\citep{Breitenbuecher2016_Diss}.
The first entities to be defined are the components forming an application, as well as the component types allowing to distinguish them and giving them semantics.
\begin{define}[Component]\label{def:component}
A \emph{component} is a physical, functional, or logical unit of an application.
\end{define}
\begin{define}[Component Type]\label{def:component-type}
A \emph{component type} is a reusable entity that specifies the semantics of a component that has this type assigned.
\end{define}
For example, a deployment model implementing the reference scenario contains several \emph{components} (\ie \emph{Order App}, \emph{Tomcat}, \emph{JSM 1.1 Queue}, \emph{Order Worker}, \emph{Ubuntu LTS}, or \emph{Azure Cosmos DB}), of different \emph{component types} (\eg \emph{Order App} is a Java-based web application, while \emph{Tomcat} is a Tomcat server).
The semantics and actions required to install or terminate a component are provided by its type.
While the component represents a certain functionality for a specific application, the component type can be used in different deployment models.
To work as intended or to provide a higher level service, components often depends on other components.
This is specified by relations between components.

\begin{define}[Relation]\label{def:relation}
A \emph{relation} is a directed physical, functional, or logical dependency between exactly two components.
\end{define}
\begin{define}[Relation Type]\label{def:relation-type}
A \emph{relation type} is a reusable entity that specifies the semantics of a relation that has this type assigned.
\end{define}
Concrete typed relations are also proposed by \cite{Weerasiri2017} and examples of them can be found in our reference scenario (cf.~\cref{fig:reference-scenario}).
For instance, the relation from \emph{Order App} to \emph{JMS 1.1 Queue} is of type \emph{connects to}, and it specifies that a network connection is to be set to allow the two components to communicate.
The relation from \emph{Order App} to \emph{Tomcat} of type \emph{hosted on}, and it indicates that the \emph{Order App} is to be installed on the \emph{Tomcat} server.

The actions and information required to realize the installation or termination of components and relations must be provided.
The EDMM encapsulate this in operations and properties.

\begin{define}[Operation]\label{def:operation}
An \emph{operation} is an executable procedure performed to manage a component or relation described in the deployment model.
\end{define}
\begin{define}[Property]\label{def:property}
A \emph{property} describes the current state or prescribes the desired target state or configuration of a component or relation.
\end{define}
Consider, for instance, the \emph{Tomcat} component in our reference application (cf.~\cref{fig:reference-scenario}) which represents a Tomcat server installed on an Ubuntu VM.
It can be associated with a property indicating that the exposed port of the web server must be \texttt{8080}.
Further, the \emph{Tomcat} component is also associated with the operation \emph{install} (cf.~\texttt{install.sh} script) that denotes the logic required to install the component on the corresponding VM.
Components and operations are implemented through so-called artifacts.
According to UML, an artifact is a physical piece of information that is created to be used for deployment and operation of a system~\citep{UML_Specification}.
On the one hand, there are artifacts representing an executable entity that implements an operation, \eg a file containing the logic required to install and start a certain application component. 
On the other hand, there are artifacts used for the operation of a component to carry out the business logic and intended functionality, \ie in the form container images, compiled binaries, or compressed source code. 
\begin{define}[Artifact]\label{def:artifact}
An \emph{artifact} implements a component or operation and are required for their execution.
\end{define}
For instance, consider the \emph{Admin App} component in the reference scenario (cf.~\cref{fig:reference-scenario}).
The shell script to install the corresponding component represents an artifact that is associated with the \emph{install} operation.
In contrast, the compiled binary file of the \emph{Admin App} also represents an artifact but this file is needed to materialize an instance of this component and is required for the operation.

Finally, note that all model entities, such as components, relations, their types as well as properties and operations are contained in a so-called deployment model.
Component and relation types are usually defined across deployment models to enable reuse including their operations as artifacts, whereas artifacts implementing a component are typically referenced in the deployment model itself.
Further, deployment models define properties, which can be referenced and used by contained elements. 
A deployment model does not have to consist of a single file, instead, it can be a set of files, which semantically belong together.
For example, several technologies allow to reference or import component types from different sources, others require a self-contained deployment model.
At the time of deployment, all entities must be available for the underlying runtime that are referenced or contained in the deployment model.
\begin{define}[Deployment Model]\label{def:deployment-model}
A \emph{deployment model} describes declaratively the desired target state of an application including all necessary model entities.
\end{define}
The target state is the completed deployment of all components and relations according to the specified properties, as described in the deployment model.

The formalized essential entities of a declarative deployment model are not only resulting from our analysis.
Similar elements have already been discussed in other research studies~\citep{Andrikopoulos2014,Breitenbuecher2016_Diss,Brogi2014_tosca-in-a-nutshell}, even if not based on a systematic review of existing and well-established declarative deployment automation technologies.

\pagebreak


\section{EDMM Technology Mapping}\label{sec:mapping}

In this section we show how the concepts of EDMM can be semantically mapped to the selected deployment technologies.
The mapping is structured into three subsections according to the categories presented in \cref{sec:categorization}.

\subsection{EDMM to GP-Technology Mapping}

We hereby show how the EDMM can be semantically mapped to
Puppet, Chef, Ansible, OpenStack Heat, Terraform, SaltStack, Juju, and Cloudify.
All of these technologies support the required features and mechanisms as outlined in \cref{subsec:analysis}.


\subsubsection{Mapping to Puppet}

In Puppet, \emph{resources} are the main building blocks describing certain aspects of the system.
For example, in our reference application, the deployment of a component can be described by a \emph{resource}.
The semantic and structure of a resource is defined by its assigned resource type (cf.~\cref{def:component-type}).
Puppet provides a set of built-in resource types which can be extended by custom resource types written in Ruby.
Resources can be bundled to \emph{modules} enabling the encapsulation of logical parts into reusable entities (cf.~\cref{def:component-type}), \eg the stack of the \emph{Order App} in our reference scenario can be encapsulated using a \emph{module}.
Resources or modules can be used in the resulting \emph{deployment model} and can be mapped to \emph{components} encompassing certain semantics.
Both, modules and resources utilize \emph{properties}, \eg to define a port a web server is listening on.
In addition, \emph{artifacts} can be defined that implement the component, \eg by defining in a module to use a compressed version of the Tomcat web server.
Modules contain a set of \emph{classes} expressing the logic to converge components into a certain state and can be mapped to \emph{operations} in EDMM.
In Puppet, \emph{relations} between components can be expressed on different levels and are limited to a predefined set of \emph{relation types}.
By including another module the semantic of a \enquote{depends on} relation type can be expressed. 
On the level of classes, a \enquote{depends on} relation can be defined by using a predefined \texttt{require} function.



\subsubsection {Mapping to Chef}

Chef uses \emph{cookbooks} to structure and encapsulate logical parts of a \emph{deployment model}.
Cookbooks can be mapped to \emph{components} as well as to \emph{component types} in our metamodel.
Using cookbooks, reusable entities can be defined expressing certain semantics, which map to \emph{component types}.
Further, cookbooks can be imported into other cookbooks, which maps semantically to \emph{components} in this context.
Our reference application (\cref{fig:reference-scenario}) can be expressed in a single cookbook by importing existing ones from the Chef supermarket, \eg a cookbook to install a Tomcat web server.
In Chef, all kinds of \emph{artifact} are supported as long as they can be referenced and packaged using the provided mechanisms.
The actual \textit{operations} to be executed to install or configure a component are encapsulated in so called \emph{recipes} written in Ruby.
Cookbooks can have dependencies to other cookbooks by using the \texttt{depends} attribute in the respective meta data file.
With \texttt{include\_recipe}, operations from dependent cookbooks can be integrated in the sequence of operations required to reach a desired state.
Thus, \emph{relations} of one \emph{relation type} can be expressed. 
By using an attribute file that, for example, defines the desired port of a web server, inputs for \emph{properties} in the recipes can be represented in Chef.



\subsubsection{Mapping to Ansible}

Ansible and Chef are similar in their concepts.
Like in Chef, playbooks are the elements used to create \emph{deployment models}.
For example, the complete reference scenario can be expressed in a single playbook. 
As playbooks can also encapsulate logical and reusable parts, they can be mapped to \emph{components} and \emph{component types} due to their recursive aggregation behavior.
There can be a generic \enquote{Tomcat} playbook that enables the deployment and configuration of an Tomcat web server.
Playbooks can also define \emph{variables} that are mapped to \emph{properties} in EDMM.
In Ansible, \emph{relations}, such as a web application is hosted on a web server, are implicitly defined by importing other Playbooks and can only expressed a \enquote{depends on} \emph{relation type}.
Ansible uses the concept of \enquote{roles}, which contain \enquote{tasks} to converge a system to a desired state.
Roles are part of playbooks and mapped to \emph{operations} and, therein, all kinds of \emph{artifacts} are supported that implement a component.



\subsubsection{Mapping to OpenStack Heat}

The \emph{deployment model} for OpenStack Heat is called Heat Orchestration Template (HOT).
The logical parts of an application, \ie its components are modeled as \emph{resources}. 
Several resource types are provided by Heat and further plugins for other resources are already available (\eg Docker or AWS) or can be created.
These resource types are reusable entities that specify the \emph{properties} and \emph{operations} that can be executed on a resource of this type.
They form the \emph{component types} in Heat.
For deployments on a VM (such as for the Admin App in \cref{fig:reference-scenario}) an infrastructure, a \texttt{Heat::SoftwareConfig}, and a \texttt{Heat::SoftwareDeployment} resource are used.
The supported \emph{artifacts} that implement a component depend on the resource type.
With a \texttt{SoftwareConfig} resource (restricted to be used only with other IaaS resources) arbitrary \textit{operations} can be specified and linked to \textit{implementation artifact}, \eg in the form of executable scripts.
To express dependencies between components, the \texttt{dependsOn} attribute can be used where one or more other components can be referenced.
Further types of \textit{relations} and, thus, \textit{relation types} can be expressed using certain properties of a component.

\subsubsection{Mapping to Terraform}

In Terraform, \emph{resources} are used to describe elements of a \emph{deployment model}, \eg compute instances, virtual networks, or software components.
Each resource is assigned to a \emph{resource type} that determines the kind of element that is managed and specifies \emph{properties} in the form of so-called \emph{attributes}.
Several resource types are provided by Terraform and custom resource types can be written in Go. 
\emph{Resources} and \emph{resource types} are mapped to \emph{components} and \emph{component types} respectively.
The supported \emph{artifacts} that implement a component depend on the \emph{resource type}.
In addition, \emph{provisioners} can be defined that are executed as part of the creation or destruction of a resource.
For example, the \emph{remote-exec provisioner} can be used to define arbitrary \emph{operations} that are executed on a resource after its creation, \eg downloading and installing an Apache Tomcat server on a provisioned EC2 instance. 
Explicit dependencies between resources can be expressed by the use of the \emph{depends\_on} attribute.
Further, \emph{modules} can be used to create logical, reusable groups of resources. 
\emph{Input Variables} on modules are mapped to \emph{properties} and are used to parameterize and customize modules. 
For example, each stack of the reference scenario can be expressed by a Terraform module specifying the required resources. 
Hence, \emph{modules} are mapped to \emph{components} and \emph{component types} in EDMM.



\subsubsection{Mapping to SaltStack}

SaltStack is a flexible orchestration and configuration management tool, which uses so-called \emph{top files} to create a \emph{deployment model}.
Top files contain \emph{formulas} and \emph{states}.
Formulas are independent and reusable entities and map to \emph{component types} in EDMM.
For example, one can create a Tomcat formula that encapsulates the logic to install and start a Tomcat web server.
Further, a formula can define a set of configuration values that map to \emph{properties} in EDMM.
Moreover, a logical group of states defined in formulas are expressing \emph{operations} which in turn relate to \emph{artifacts} that implement these.
Using their DSL syntax, arbitrary logic in the form of states can be supplied to reach a desired state.
By using a formula in a deployment model a \emph{component} is created according to our metamodel.
\emph{Relations} in SaltStack are, on the one hand, derived by the sequence of used formulas in a deployment model.
On the other hand, states can \emph{depend on} other stats defined by certain formulas, which results in a certain execution order.


\subsubsection{Mapping to Juju}

Juju is a topology-based application modeling tool based on a declarative YAML DSL.
All instructions and artifacts necessary for deploying and configuring application components are defined in \emph{charms}, which map to \emph{component types} in the EDMM as they are reusable entities having a certain semantic.
Each charm provides a set of configuration values that can be set during deployment, which are mapped to \emph{properties}.
Further, a charm defines actions, implemented as scripts, that are triggered by the runtime during deployment.
These are respectively mapped to \emph{operations} and \emph{artifacts}, whereas artifacts that implement components are carried out by these operations.
Charms can be used in \emph{bundles}, which implies that a \emph{component} of a certain \emph{component type} inside a \emph{deployment model} is used.
In Juju, there can be \emph{relations} following a \enquote{depends on} semantic by expressing requirements and capabilities on charms.
For example, charm defines that it requires a database and, correspondingly, a database charm is capable of satisfying this requirements.
This can be expressed using the \texttt{relations} keyword in the model.
A compound deployment including multiple charms, their configuration, and relations can be described in a Juju bundle.



\subsubsection{Mapping to Cloudify}

The DSL defined and used by Cloudify is based on the TOSCA YAML profile~\citep{TOSCA_SimpleProfile}.
However, the standard is not completely met.
Similar to OpenStack Heat, built-in types encompassing Cloudify basic types can be used to model \emph{components}.
Further \emph{components types} can be made available using plugins.
In Cloudify, \emph{node types} and \emph{node templates} are mapped to \emph{component types} and \emph{components} respectively according to the EDMM definitions.
Using Cloudify's lifecycle interface, \ie \emph{operations} allow to create, start, stop, and terminate physical resources.
Node types define \emph{properties}, implement operations, and define deployment as well as implementation \emph{artifacts}.
Built-in relation types, for example, define a \texttt{depends\_on} or \texttt{connected\_to} relation between components.
In contrast to all other considered technologies, further relation types can be defined.
To realize the reference scenario from \cref{fig:reference-scenario}, the AWS and Azure plugins are required.
These plugins provide all types, operations, and artifacts to model the required components as well as to interact with the respective cloud provider application-programming-interfaces (API).

\subsection{EDMM to ProvS-Technology Mapping}

We hereby show how the EDMM can be semantically mapped to
AWS CloudFormation and Azure Resource Manager.
In contrast to GP, these technologies only support single-cloud deployments as they only support the services of the respective cloud provider.


\subsubsection{Mapping to AWS CloudFormation}

AWS CloudFormation is the deployment and management service by AWS and uses a JSON or YAML templates to create \emph{deployment models}.
In the template, \emph{resources} are used to express \emph{components}.
AWS provides a set of built-in \emph{resource types}, referred as \emph{component types}, that specify the semantics of components, \eg defining \emph{properties} supported by a resource.
CloudFormation enables to create reusable \emph{component types} by defining \emph{stacks} that can be in turn used in other templates.
Each stack of the reference scenario in \cref{fig:reference-scenario} can be modeled by one or more resources.
For example, the AWS SQS and Lambda service an be used to implement the depicted \emph{JMS 1.1 Queue} and \emph{Order Worker} components. 
To deploy the \emph{Admin App} an AWS EC2 instance can be defined where one can specify the required installation and configuration steps as an \emph{operations}, provided in separate files.
The semantic of relations is restricted as only one type of inter-component dependency can be specified, by using the attribute \texttt{dependsOn} on resources.

\subsubsection{Mapping to Azure Resource Manager}

In Azure Resource Manager (ARM), JSON templates describe the configuration of Azure resources and services. 
Azure services (\eg compute instances, databases, or middleware) are modeled as \emph{resources}.
The structure and semantics of a resource (\eg its supported \emph{properties} and \emph{artifacts}) are defined by built-in \emph{resource types}. 
Hence, \emph{resource} and \emph{resource type} map to \emph{component} and \emph{component type} in the EDMM.
For example, the MongoDB configuration hosted on Azure Cosmos DB can be expressed using a resource of type \texttt{Microsoft.DocumentDB/databaseAccounts}.
\emph{Relations} between resources can be specified using the \texttt{dependsOn} element defining a dependency to one or more resources.
The resources of a deployment can either be defined in a single template or divided into multiple ones in order to create purpose-specific, reusable templates.
As ARM templates are logical and reusable units, they are mapped to \emph{components} and \emph{component types} in the EDMM.  
Post-deployment configurations, software installations, or other actions to configure a VM, can be achieved through \emph{virtual machine extensions}, which are semantically mapped to \emph{operations} in the EDMM.

\subsection{EDMM to PlatS-Technology Mapping}

We hereby show how the EDMM can be semantically mapped to
Kubernetes, CFEngine, and Docker Compose.
In contrast to GP, these technologies are restricted to specific services (XaaS support) and to the use of specific \emph{platform bundles} for realizing components.


\subsubsection{Mapping to Kubernetes}

With Kubernetes, developers can specify the \emph{deployment model} of a multi-service application by indicating the \enquote{pods} to run, one for each \emph{service} forming the application.
Their desired configuration can then be specified by defining \enquote{deployments}, each targeting a different subset of pods.
For instance, each component in our reference application (cf.~\cref{fig:reference-scenario}) should be placed in a different pod, and its desired configuration could be specified by defining a different deployment targeting its corresponding pod. 
Kubernetes hence provides a predefined set of \emph{component types} allowing to define reusable units of pods, deployments, and services (\emph{components} of an application).
Kubernetes also supports a predefined set of \emph{relation types} in the form of specifying which pods are targeted by which deployment or service.
Attributes specifying the desired configuration for pods, deployments, and services are mapped to \emph{properties} in the EDMM.
In Kubernetes, \emph{artifacts} for implementing \emph{components} are reflected by container images, which contain the complete stack starting from the operating system to the application-specific component, depending on application requirements~\citep{Pahl2017_cloud-container-technologies}.
Moreover, container images also encapsulate \emph{operations} representing the logic to install and configure the components. 
Therefore, container images are \emph{platform bundles} as they are the unit of deployment.



\subsubsection{Mapping to CFEngine}

CFEngine assumes an already running computing infrastructure and, therefore, is assigned to the PlatS deployment technologies. 
In CFEngine, everything is a \emph{promise}.
Promises are used to define the desired state that should be reached, \eg a package to be installed or a process to be started.
Further, \emph{bundles} can be used to logically group promises and are, therefore, mapped to \emph{operations} in EDMM.
\emph{Bodies} are used to create reusable parts of promises and are mapped to \emph{component types}.
Bodies can also define \emph{properties} and once they are used in a \emph{deployment model} a concrete \emph{component} is created.
There can be explicit \emph{relations} between components that specify the required execution order using the \texttt{depends\_on} property on promises.



\subsubsection{Mapping to Docker Compose}

Docker Compose permits specifying the \emph{deployment model} of a multi-container Docker application in a single file. 
The file is organized in \enquote{services}, which are used to specify the \emph{components} (\ie its containers) of an application.
\emph{Relations} between components are expressed using the \texttt{depends\_on} keyword.
Mapping this to our metamodel, the only \emph{component type} and \emph{relation type} are expressed by \enquote{services} and by the \texttt{depends\_on} attribute, respectively.
Docker Compose predefines the set of \emph{properties} that can be associated with the services forming an application.
\emph{Properties} can be associated to each service in a Docker Compose file and specify the desired configuration.
\emph{Artifacts} and \emph{operations} are a special case since they have to be packaged as a so-called \emph{platform bundle}.
The artifact implementing a component as well as the logic to install and configure it must be supplied through a Docker image, either retrieved from a repository or built based on a \texttt{Dockerfile}. 
For example, to deploy the \emph{Order App} the complete stack starting from the operating system to the application-specific component must be linked into a Docker image.

\subsection{EDMM to TOSCA}

As various technologies support the TOSCA standard, \ie OpenTOSCA~\citep{Binz2013_OpenTOSCA}, ALIEN 4 Cloud \citep{Alien4Cloud2018}, Cloudify, and TosKer~\citep{Brogi2018_TosKer}, this section presents a mapping of EDMM to TOSCA---although it was out of scope of this paper due to its rank.
EDMM only uses a subset of entities specified by the standard:
\emph{Service templates} are used to express \emph{deployment models}, while \emph{components} and \emph{component types} are referred as \emph{node templates} and \emph{node types}, respectively.
TOSCA allows the definition of arbitrary \emph{relations} and \emph{relation types}, called \emph{relationship template} and \emph{relationship type}, but defines a certain set of normative types every compliant orchestrator needs to support, \ie \texttt{hosted\_on} and \texttt{connected\_to}.
Node types and relationship types support the definition of \emph{properties} as they are used to define semantics.
Further, \emph{operations} in EDMM are mapped to \emph{management operations}, which are realized by \emph{implementation artifacts}.
In contrast, there are \emph{deployment artifacts} that are mapped to \emph{artifacts} required for the execution of a component.

\pagebreak


\section{Related Work}\label{sec:relatedwork}

In this section we discuss closely related work on reviews and comparisons regarding cloud computing deployment technologies and their respective meta modeling results.
\cite{Weerasiri2017} introduced a taxonomy for cloud resource orchestration based on a survey examining eleven cloud orchestration approaches.
They describe the notion of a \emph{Resource Entity Model} consisting of entities, relationships, and constraints.
The Resource Entity Model shows on a high level the structure of a cloud application.
However, to automate a deployment more information are required, \eg what artifacts have to be installed and how. 
Therefore, we have examined the semantics of used deployment technologies and formulated a metamodel containing the common elements.

A detailed comparison of six different \emph{Infrastructure-as-Code} platforms has been conducted by \cite{Masek2018_AnsibleFramework}.
They distinguish between \enquote{configuration management} tools, designed to install and manage software on existing nodes, and \enquote{orchestration tools}, designed to provision the servers themselves and leaving the job of configuring nodes to other tools.
Also, \cite{Wettinger2015_DevOps_TOSCA} introduce a similar classification by differentiating between node-centric and environment-centric artifacts.
Node-centric artifacts are deployment models that are executed on single nodes, such as Chef or Ansible, whereas environment-centric artifacts are deployment models that are executed on a higher level including more that one node, such as Terraform.
In both works, the essence is that these two categories are not mutually exclusive.
Most configuration management tools can do some degree of provisioning and most orchestration tools can do some degree of configuration management or can even integrate other tools.
We derived a different grouping criteria resulting from our review in order to make a clear assignment for each technology.

Besides the variety of tools, there are also standards in the field of application deployment.
\cite{Markoska2015} give an brief overview about different cloud deployment technologies and standards, such as AWS CloudFormation, TOSCA, CAMP, and others.
\cite{Martino2015} focuses only on the TOSCA standard and  OpenStack Heat templates for a qualitative comparison.
Further, \cite{Bergmayr2018_CMLReviewJournal} provide a very detailed overview and comparison of different cloud modeling languages (CML).
They conducted a systematic review of CMLs, their features, and discuss core domain concepts of such.
However, these studies do not review a variety of used deployment automation technologies and, further, do not abstract to a commonly denominated metamodel.

\noindent
\cite{VergaraVargas2017} developed a new Architecture Description Language (ADL) that comprises deployment aspects.
One part of their ADL is to support software deployments based on a model-driven deployment (MDDep) approach expressing components and relations among them.
\cite{Alipour2018} focuses on a model-driven approach presenting a \emph{Cloud Platform Independent Model} in order to deploy auto-scaling services in a cloud-agnostic way.
In general literature about software architectures (SAs), SAs are described as structures that comprise software elements, relations among them, and properties of both~\citep{Rozanski2012}.
Also \cite{Chen1976} describes in his paper a unified view to model data, consisting of entities, and relationships.
Thus, these elements are similar to the parts identified by the EDMM.
Further, representing an application structure as a graph is a common approach in research.
For example, GENTL~\citep{Andrikopoulos2014} is a CML to express topologies whereas \cite{Breitenbuecher2016_Diss} proposes a graph-based description language (DMMN) to enable the declarative modeling of management activities~\citep{Breitenbuecher2016_Diss}.
However, as they are proposing similar elements, these findings are not based on the analysis of widely used industrial tools.




\section{Conclusions and Outlook}\label{sec:conclusion}


We conclude that there are three EDMM groups a technology can be assigned to: 
General-purpose (GP) technologies support multi-cloud and XaaS deployments, provider-specific (ProvS) technologies are restricted to the respective cloud provider services, and platform-specific (PlatS) technologies support only a subset of XaaS and require specific platform bundles for deployment.
A understanding of essential deployment model elements helps to compare technologies regarding deployment features and mechanism and supports decision making processes when selecting an appropriate technology for an use case.
The introduced classification and the presented EDMM technology mapping support the migration from one deployment technology into another one.
Further, this does not only support industry to compare and select technologies, but also helps researcher to evaluate concepts in the area of deployment automation research: 
If new research can be realized using EDMM, our mappings prove that this research can be also applied to the technologies analyzed in this paper. 
This significantly eases practically validating new concepts.






\begin{acknowledgements}
  This work is partially funded by the European Union's Horizon 2020 research and innovation project \emph{RADON} (825040), the BMWi projects \textit{IC4F} (01MA17008G) and \textit{SePiA.Pro} (01MD16013F), the DFG project ADDCompliance (636503), the POR-FSE project \textit{AMaCA}, and the project \textit{DECLware} (PRA\_2018\_66, University of Pisa).
\end{acknowledgements}

\pagebreak

\bibliographystyle{spbasic} 
\bibliography{paper} 


\end{document}